\documentclass{SciPost}

\usepackage{array}
\usepackage{listings}
\usepackage{xcolor}
\usepackage{orcidlink}

\definecolor{comment-color}{rgb}{0.8,0.1,0.1}
\definecolor{keyword-color}{rgb}{0.2,0.2,1}
\definecolor{string-color}{rgb}{0.5, 0, 0.8}
\definecolor{bg-gray}{gray}{0.90}
\definecolor{codegray}{gray}{0.90}

\lstdefinestyle{mypython}{
  language=python,
  basicstyle=\ttfamily\footnotesize,
  stringstyle=\color{string-color},
  keywordstyle=\color{keyword-color},
  commentstyle=\color{comment-color}\itshape\footnotesize,
  fontadjust=true,
  morekeywords={np,plt,SimpleED,Fragment,Lattice},
  morecomment=[l]{\#\ },
  mathescape,
  fontadjust=true,
  mathescape,
  breakatwhitespace=false,
  keepspaces=true,
  showstringspaces=false,
  columns=fullflexible,
  backgroundcolor=\color{bg-gray},
  numbers=none, numberstyle=\tiny, stepnumber=1, numbersep=3pt
}

\lstdefinestyle{mybash}{
  language=bash,
  basicstyle=\ttfamily\footnotesize,
  stringstyle=\ttfamily,
  keywordstyle=\color{keyword-color},
  commentstyle=\color{comment-color}\itshape\footnotesize,
  morekeywords={git,mkdir,cmake,make},
  alsoletter=-
  captionpos=b,
  mathescape,
  fontadjust=true,
  breakatwhitespace=false,
  keepspaces=true,
  showstringspaces=false,
  columns=flexible,
  xleftmargin=3.4pt,
  xrightmargin=3.4pt,
  backgroundcolor=\color{bg-gray}, 
  frame=none
}

\hypersetup{
    colorlinks,
    linkcolor={red!50!black},
    citecolor={blue!50!black},
    urlcolor={blue!80!black}
}

\usepackage[bitstream-charter]{mathdesign}
\DeclareSymbolFont{usualmathcal}{OMS}{cmsy}{m}{n}
\DeclareSymbolFontAlphabet{\mathcal}{usualmathcal}

\fancypagestyle{SPstyle}{
\fancyhf{}
\lhead{\colorbox{scipostblue}{\bf \color{white} ~SciPost Physics Codebases }}
\rhead{{\bf \color{scipostdeepblue} ~Submission }}

\fancyfoot[C]{\textbf{\thepage}}
}

\begin{document}

\pagestyle{SPstyle}

\begin{center}{\Large \textbf{\color{scipostdeepblue}{
GEM: An implementation of the ghost-Gutzwiller approximation for simulating interacting quantum systems
}}}\end{center}

\begin{center}\textbf{
Samuele Giuli\,\orcidlink{0009-0004-7341-3655}\textsuperscript{1$\star$},
Tsung-Han Lee\,\orcidlink{0000-0002-0571-9909}\textsuperscript{2,3},
Yong-Xin Yao\,\orcidlink{0000-0002-7830-5942}\textsuperscript{4,5},
Ina Park\,\orcidlink{0000-0003-0161-711X}\textsuperscript{1},
Harrison LaBollita\,\orcidlink{0000-0002-6699-8577}\textsuperscript{1},
Ivan Pasqua\,\orcidlink{0009-0004-7603-8037}\textsuperscript{1},
Nils Wentzell\,\orcidlink{0000-0003-3613-007X}\textsuperscript{1},
Nicola Lanatà\,\orcidlink{0000-0003-0003-4908}\textsuperscript{1,6} and
Olivier Gingras\,\orcidlink{0000-0003-3970-6273}\textsuperscript{1,7$\dagger$}
}
\end{center}

\begin{center}
{\bf 1} Center for Computational Quantum Physics, Flatiron Institute, 162 Fifth Avenue, New York, New York 10010, USA
\\
{\bf 2} Department of Physics, National Chung Cheng University, Chiayi 62102, Taiwan
\\
{\bf 3} Physics Division, National Center for Theoretical Sciences, Taipei 10617, Taiwan

{\bf 4} Ames National Laboratory, Ames, Iowa 50011, USA
\\
{\bf 5} Department of Physics and Astronomy, Iowa State University, Ames, Iowa 50011, USA
\\
{\bf 6} School of Physics and Astronomy, Rochester Institute of Technology, 84 Lomb Memorial Drive, Rochester, New York 14623, USA
\\
{\bf 7} Université Paris-Saclay, CNRS, CEA, Institut de physique théorique, 91191, Gif-sur-Yvette, France
\\[\baselineskip]
$\star$ \href{mailto:sgiuli@flatironinstitute.org}{\small sgiuli@flatironinstitute.org}\,,\quad
$\dagger$ \href{mailto:ogingras@flatironinstitute.org}{\small ogingras@flatironinstitute.org}
\end{center}

\section*{\color{scipostdeepblue}{Abstract}}
\textbf{\boldmath{%
We present \textsc{GEM} (Ghost Embedding Method), an open-source software package written in Python for computing equilibrium properties of strongly correlated electronic systems within the ghost-Gutzwiller approximation method.
GEM provides a computationally efficient framework for studying multi-orbital lattice models.
It supports zero- and finite-temperature calculations and symmetry broken phases.
It is integrated with the \textsc{TRIQS} ecosystem, providing tools for model construction, self-consistent solution, and evaluation of physical observables.
We first detail the method's theoretical formulation, then we present the software architecture, and finally we introduce some practical workflow, which also validates the implementation against established results.
In particular, we illustrate the capabilities of GEM through multiorbital and finite-temperature applications and discuss its computational cost relative to more demanding quantum embedding approaches.

}}

\vspace{\baselineskip}

\noindent\textcolor{white!90!black}{%
\fbox{\parbox{0.975\linewidth}{%
\textcolor{white!40!black}{\begin{tabular}{lr}%
  \begin{minipage}{0.6\textwidth}%
    {\small Copyright attribution to authors. \newline
    This work is a submission to SciPost Physics Codebases. \newline
    License information to appear upon publication. \newline
    Publication information to appear upon publication.}
  \end{minipage} & \begin{minipage}{0.4\textwidth}
    {\small Received Date \newline Accepted Date \newline Published Date}%
  \end{minipage}
\end{tabular}}
}}
}


\vspace{10pt}
\noindent\rule{\textwidth}{1pt}
\tableofcontents
\noindent\rule{\textwidth}{1pt}
\vspace{10pt}


\section{Introduction}
\label{sec:Introduction}

Nonperturbative calculations for correlated-electron models require a compromise between physical fidelity, numerical cost, and the range of accessible observables.
Quantum embedding methods provide one of the main routes for treating strong local correlations while retaining a tractable computational problem.
Among those, dynamical mean-field theory (DMFT)~\cite{Georges1996} played a central role in the understanding of the Mott transition and the effects of strong correlations.
Its self-consistency loop requires repeated solutions of an interacting quantum impurity problem and the evaluation of frequency-dependent quantities, namely the impurity Green’s function.
In most cases of interest, obtaining these solutions quickly becomes numerically prohibitive as one increases the amount of orbitals, lower the temperature, wants to perform broad parameter scans or is interested in symmetry-broken phases.

The conventional Gutzwiller approximation (GA) is a substantially lighter approach which provides a good description of the renormalized low-energy quasiparticles but does not reproduce the incoherent high-energy Hubbard bands~\cite{Gutzwiller1965}.
One of its extension, the ghost-Gutzwiller approximation (ghost-GA)~\cite{Lanata2017}, consists of augmenting the variational auxiliary space in order to recover high-energy features in the spectral function and to reach DMFT-like total energies and static expectation values already with a moderate enlargement of the auxiliary space~\cite{Lanata2017,Lee2023}.
Recently, an approach based on a free-energy functional has proven the equivalence between DMFT and ghost-GA in the limit of infinite auxiliary states, providing a route to extend ghost-GA to perform finite temperature calculations~\cite{Giuli2026}, yet at the moment, no open-source software provides an implementation of this formulation.

GEM addresses this gap by making the systematically improvable ghost-GA formulation, both at zero- and finite-temperature, available within an open-source research software.
The implementation follows the free-energy functional that unifies zero-temperature ghost-GA, its finite-temperature extension, and DMFT: stationary conditions are expressed using static or thermal expectation values of effective Hamiltonians. DMFT is obtained as the number of auxiliary bath modes per physical orbital, $B$, tends to infinity: $B\rightarrow\infty$~\cite{Giuli2026}.
Although the exact correspondence holds in the $B \rightarrow \infty$ limit, convergence can be rapid in practice~\cite{Lee2023}. For the finite-temperature Hubbard-model benchmarks reported in Ref.~\cite{Giuli2026}, calculations with only $B=3$ already reproduce DMFT thermodynamic results with high accuracy across the investigated interaction and temperature regimes.
Moreover, already the much simpler GA (ghost-GA with $B=1$) gives a good description of the thermodynamic stability of symmetry broken phases, and provides a good estimate for total energies and other static observables like order parameters~\cite{smit2025,KazemiMoridani2026,Bellomia2026}.

The software ecosystem for quantum embedding is by now well developed, particularly for DMFT, where mature implementations exist for a broad range of impurity solvers and many-body workflows.
These include exact-diagonalization~\cite{Caffarel1994,Amaricci2022,Crippa2025,SciPostPhysCodeb.23}, continuous-time quantum Monte Carlo~\cite{Werner2006_1,Werner2006_2,PhysRevB.75.155113, Gull2011, CTHYB2016, Wallerberger2019, MELNICK2021108075, CTSEG2025}, tensor-network~\cite{Wolf2015, Ganahl2015}, and numerical renormalization group approaches~\cite{Zitko2009}, together with broader frameworks such as \textsc{triqs}~\cite{Parcollet2015} and \textsc{alps}~\cite{Bauer2011} that provide reusable infrastructure for Green's functions, operators, data handling, and solver interoperability.
This maturity makes DMFT a flexible and powerful tool, but many applications do not require the full cost and complexity of a fully dynamical impurity calculation.

Gutzwiller-based approaches provide a complementary route in this regime.
Publicly available implementations of the Gutzwiller approximation and ghost-GA exist, such as CyGutz~\cite{Yao2015} and Portobello~\cite{ADLER2024108907}, but no maintained open-source package implements the finite-temperature ghost-GA formulation recently described in Ref.~\cite{Giuli2026}.
GEM addresses this gap by providing a variational embedding hierarchy controlled by the number $B$ of auxiliary bath modes (ghost orbitals).
Increasing $B$ systematically enlarges the auxiliary space and approaches the DMFT limit, while small values of $B$ can already provide an advantageous balance between computational cost and accuracy~\cite{Giuli2026}.
GEM therefore occupies an intermediate regime between inexpensive static Gutzwiller approximations and fully dynamical DMFT workflows, enabling ground-state and thermodynamic calculations without requiring a full frequency-dependent impurity solution at every stage.

As a result, GEM adopts a package design that implements the ghost-embedding algorithm and its core data structures, while remaining interoperable with external numerical and impurity-solver components.
At the moment, it comes with an internal ED impurity solver, but it also offers a template to generate external impurity solvers.
In the future, impurity solvers based on various numerical methods will be interfaced, providing seamless access to a variety of options depending on the physical systems of interest.
This positioning of GEM enables two complementary workflows: 1)~It can be used as a lower-cost embedding method for static and thermal observables at moderate auxiliary-space size, for which it has been shown that the agreement with DMFT is already excellent at small bath sizes~\cite{Lee2023}, and 2)~converged ghost-GA solutions can be used as a starting point for DMFT as shown in Ref.~\cite{makaresz2026}.

The target users are researchers developing or applying quantum embedding methods in condensed matter physics and materials theory.
Representative uses include the study of Hubbard models and interacting multiorbital systems, phase-diagram calculations, tests of convergence with auxiliary-space size, investigations of Mott and orbital-selective transitions, and benchmarking or developing impurity solvers.
By providing the method as a reusable package rather than a collection of project-specific scripts, GEM supports reproducible comparisons across models, solvers, temperatures, and approximation levels.

This manuscript is structured as follows. In Sec.~\ref{sec:Installation}, we provide information on how to install the software. In Sec.~\ref{sec:Method}, we give a brief description of the method, followed in Sec.~\ref{sec:Implementation} by a description of its implementation in GEM. In Sec.~\ref{sec:examples}, we present three examples of prototypical calculations illustrating the different capabilities of the software: 1)~finite temperature calculations of a single-orbital model for total energy and free energy, 2)~finite temperature calculations of a single-orbital model for the phase diagrams of symmetry broken phases, and 3)~zero temperature calculations of a multi-orbital Hund system for various interaction strengths. In Sec.~\ref{sec:outlook}, we provide an outlook on future directions for the development of the software and in Sec.~\ref{sec:conclusions}, we summarize the content of the manuscript.

\section{Installation}
\label{sec:Installation}

Since GEM is pure Python, building and installing it requires no additional compilation. Its only dependencies are shown in Table~\ref{tab:dependencies}. These dependencies are a
Python 3 interpreter, CMake, and four scientific Python packages: NumPy, SciPy, Numba and h5py. The versions used to test the current release are listed in Table~\ref{tab:dependencies}. These versions are known to work with GEM but should not be interpreted as minimum version requirements.
Since each Numba release supports a bounded range of NumPy versions, the two should be upgraded together.

\begin{table}[ht!]
    \centering
    \begin{tabular}{c|c|c}
        \hline
        Software & Ref. & Tested  \\
        \hline
        \hline
        Python   & \cite{van1995python}                             &   3.9.6 \\
        CMake    & \cite{Martin2015MasteringC}                      & 4.3.2   \\
        NumPy    & \cite{harris2020array}                           & 2.0.2   \\
        SciPy    & \cite{2020SciPy-NMeth}                           & 1.13.1  \\
        Numba    & \cite{10.1145/2833157.2833162}                   & 0.60.0  \\
        h5py     & \cite{The_HDF_Group_Hierarchical_Data_Format}        & 3.14.0 \\
        \hline
    \end{tabular}
    \caption{External software dependencies of GEM along with versions known to work.}
    \label{tab:dependencies}
\end{table}

Optionally, GEM offers an MPI-based parallelization scheme using mpi4py~\cite{Dalcin2021mpi4py}. Details of the parallelization are provided in Section~\ref{sec:Implementation}. If mpi4py is not installed, GEM falls back to a serial calculation, with parallelism limited to any multithreading provided by the LAPACK and BLAS libraries linked to the installed Python packages.

The internal \texttt{SimpleED} solver requires nothing beyond these packages; interfaces to external impurity solvers require their corresponding
libraries, which should be cited alongside GEM when used. Some examples additionally use Matplotlib~\cite{Hunter:2007} for plotting and \textsc{triqs}~\cite{Parcollet2015} for lattice construction.
The package is built and installed out of source with CMake:
\begin{lstlisting}[style=mybash]
git clone https://github.com/TRIQS/gem.git
mkdir gem.build && cd gem.build
cmake ../gem
make
make test
make install
\end{lstlisting}

By default the installation prefix is the one associated to the Python interpreter found by CMake—typically the interpreter of an active virtual environment—so that \texttt{import gem} works directly after installation, and the test suite is built and run by \texttt{make test}.
The configuration step accepts optional flags, passed as 
\texttt{-D<Option>=<value>} on the \texttt{cmake} line, in any number and order. The available flags as listed in Table~\ref{tab:cmake_options}.

\begin{table}[ht!]
    \centering
    \begin{tabular}{|c|>{\centering\arraybackslash}p{7.8cm}|}
        \hline
         Option & Effect \\
         \hline
         \texttt{-DCMAKE\_INSTALL\_PREFIX=<path>} &  install into \texttt{<path>} instead of the Python prefix \\
         \texttt{-DBuild\_Tests=OFF} &  skip the test suite (\texttt{ON} by default) \\
         \texttt{-DBuild\_Documentation=ON} & build documentation (\texttt{OFF} by default; requires Sphinx~\cite{sphinx}) \\\hline
    \end{tabular}
    \caption{Options that can be passed when calling \texttt{cmake}.}
    \label{tab:cmake_options}
\end{table}

\section{Method}
\label{sec:Method}

GEM implements ghost-GA in the quantum-embedding formulation introduced in Refs.~\cite{Lanata2017,Giuli2026}.
In this section, we briefly summarize the formulation used in the code, with particular emphasis on the quantities entering the numerical self-consistency.
For a more complete derivation and a discussion of the connection between ghost-GA and DMFT, we refer the reader to Ref.~\cite{Giuli2026}.

\subsection{Interacting lattice problem}
\label{ssec:lattice_problem}

We consider a general multi-orbital lattice Hamiltonian
\begin{equation}
\hat H = \hat H_0 + \hat H_{\mathrm{int}},
\end{equation}
with a local interaction
\begin{equation}
\hat H_{\mathrm{int}}
=
\sum_{i=1}^{\mathcal N}
\hat H^i_{\mathrm{int}},
\end{equation}
and a one-body contribution
\begin{equation}
\hat H_0 =
\sum_{i,j=1}^{\mathcal N}
\sum_{\alpha=1}^{\nu_i}
\sum_{\beta=1}^{\nu_j}
[h_0]_{i\alpha,j\beta}
\hat{c}^\dagger_{i\alpha}\hat{c}_{j\beta}.
\end{equation}
Here, $\hat{c}^\dagger_{i\alpha}$ ($\hat{c}_{i\alpha}$) is an operator that creates (annihilates) an electron on the correlated fragment labeled by $i$ (often site) with physical fermionic degrees of freedom labeled by $\alpha$. We have $i=1,\ldots,\mathcal{N}$ where $\mathcal{N}$ is the total number of correlated fragments, and $\alpha = 1,\ldots, \nu_i$ where $\nu_i$ is the number of spin-orbital electronic levels in the fragment $i$. $h_0$ is a matrix of scattering amplitudes.

In the present form of the code, we only consider cases for which the intra-fragment interaction is two-body and can be expressed as:
\begin{equation}
    \hat H^i_{\text{int}} = \sum_{i=1}^{\mathcal N}
\sum_{\alpha, \beta, \gamma, \delta=1}^{\nu_i} \frac{U^i_{\alpha \beta \gamma \delta}}{2} \hat{c}^\dagger_{i \alpha} \hat{c}_{i \beta} \hat{c}^\dagger_{i \gamma} \hat{c}_{i \delta},
\end{equation}
but the derivation is completely general.
It is convenient to separate the local intra-fragment one-body contribution from the
inter-fragment hopping,
\begin{equation}
h_0 = \epsilon + t,
\end{equation}
where $\epsilon=\mathrm{diag}(\epsilon_1,\ldots,\epsilon_{\mathcal N})$, or $[\epsilon]_{i\alpha,j\beta} = [h_0]_{i\alpha,i\beta} \delta_{ij} \equiv [h_0^i]_{\alpha\beta} \delta_{ij}$,
contains the local one-body blocks and $t$ contains the off-diagonal hopping
terms.
The local one-body contribution can then be combined with the interaction into
the local Hamiltonian
\begin{equation}
\hat H^i_{\mathrm{loc}}
=
\sum_{\alpha,\beta=1}^{\nu_i}
[\epsilon_i]_{\alpha\beta}
\hat{c}^\dagger_{i\alpha}\hat{c}_{i\beta}
+
\hat H^i_{\mathrm{int}}.
\label{eq:Hloc}
\end{equation}
The full Hamiltonian can therefore be written as
\begin{equation}
\hat H =
\sum_{i\neq j}
\sum_{\alpha=1}^{\nu_i}
\sum_{\beta=1}^{\nu_j}
[t_{ij}]_{\alpha\beta}
\hat{c}^\dagger_{i\alpha}\hat{c}_{j\beta}
+
\sum_i \hat H^i_{\mathrm{loc}}.
\label{eq:H_kin_and_loc}
\end{equation}

When the system is translationally invariant, with a primitive unit cell that contains one or more fragments, one can rewrite the Hamiltonian as follows:
\begin{equation}
\hat H =
\sum_{I,J}
\sum_{i \in I,j \in J}
\sum_{\alpha=1}^{\nu_i}
\sum_{\beta=1}^{\nu_j}
[t^{IJ}_{i j}]_{\alpha\beta}
\hat{c}^\dagger_{Ii \alpha}\hat{c}_{Jj \beta}
+
\sum_I \sum_{i \in I} \hat H^{Ii}_{\mathrm{loc}}.
\label{eq:H_kin_and_loc_supercell_IJ}
\end{equation}
Here, $I$ ($J$) is the index of a superlattice cell and $i$ ($j$) is index running over the fragments belonging to a superlattice cell.
Transforming the kinetic part of Eq.~\eqref{eq:H_kin_and_loc_supercell_IJ} to reciprocal space, we get the following Hamiltonian:
\begin{equation}
\hat H =
\sum_{K}
\sum_{i,j}
\sum_{\alpha=1}^{\nu_i}
\sum_{\beta=1}^{\nu_j}
[t^{K}_{i j}]_{\alpha\beta}
\hat{c}^\dagger_{Ki \alpha} \hat{c}_{Kj \beta}
+
\sum_I \sum_{i \in I} \hat H^{Ii}_{\mathrm{loc}}.
\label{eq:H_kin_and_loc_supercell_K}
\end{equation}
where $K$ is a reciprocal lattice vector belonging to the Brillouin zone of the superlattice, $t^K$ is the Fourier transformed hopping matrix and $\hat{c}^\dagger_K$ ($\hat{c}_K$) are the Fourier transformed creation (annihilation) operators.

\subsection{Dynamical functional and ghost embedding}
\label{ssec:dynfunc}

The starting point of the formulation is the finite-temperature dynamical functional
\begin{align}
\mathcal{L}_{\beta} [{\Sigma_i},{\Delta_i}]
=&
-\beta^{-1}\sum_n e^{i\omega_n0^+}
\mathrm{Tr}\ln
\big[
i\omega_n\mathbf{1}
-h_0-\Sigma(i\omega_n)
\big]
\nonumber \\
&+
\beta^{-1}\sum_i\sum_n e^{i\omega_n0^+}
\mathrm{Tr}
\ln
\big[
i\omega_n\mathbf{1}
-\epsilon_i
-\Delta_i (i\omega_n)
-\Sigma_i (i\omega_n)
\big]
\nonumber \\
&+
\sum_i
\Omega_{\mathrm{imp}}^i [\Delta_i],
\label{eq:dynamical_functional}
\end{align}
where
$\Sigma=\mathrm{diag}(\Sigma_1,\ldots,\Sigma_{\mathcal N})$
is a block matrix, $\Sigma_i$ is the self-energy of fragment $i$ which contains all the intra-fragment electronic correlations, $\Delta_i$ is the hybridization function associated with fragment $i$,
and $\Omega_{\mathrm{imp}}^i$ is the free energy of the corresponding interacting impurity problem. We work in imaginary frequencies, with $\omega_n$ the n$^{\text{th}}$ Matsubara frequency, and $\beta$ is the inverse temperature considered.
As demonstrated in Ref.~\cite{Giuli2026}, the stationary conditions of the functional in Eq.~\eqref{eq:dynamical_functional} are the DMFT self-consistency conditions~\cite{Georges1996}.

Given a positive integer $B$, which quantifies the number of ghost degrees of freedom and therefore the size of the variational space, one can introduce~\cite{Giuli2026} the following parametrization for the self-energies and hybridization functions of fragment $i$:
\begin{align}
\Sigma_i (z) &= z \mathbf{1}_{\nu_i} -
    \left[
\mathcal R_i^\dagger
\left( z\mathbf{1}_{B\nu_i}-\Lambda_i \right)^{-1}
\mathcal R_i
\right]^{-1}
- \epsilon_i ,
\label{eq:self_parametrization}
\\
\Delta_i(z)
&=
\mathcal D_i^T
\left( z\mathbf{1}_{B\nu_i}+\Lambda_i^c \right)^{-1}
\mathcal D_i^* .
\label{eq:hybr_parametrization}
\end{align}
where $\Lambda_i, \Lambda^c_i \in \mathbb{C}_{B \nu_i \times B\nu_i}$ are Hermitian matrices, $R_i^\dagger,\mathcal{D}_i \in \mathbb{C}_{B \nu_i \times \nu_i}$, and $z \in \mathbb{C}$.
Within this parametrization, the functional recover the ghost-GA functional at zero temperature and provide a natural finite temperature extension, as shown in Ref.~\cite{Giuli2026}.
Here, $B$ is a parameter that determines the enlargement of the auxiliary space with respect to the physical one. With $B=1$, the auxiliary space and the physical space have the same dimensionality and at zero temperature, one recovers the standard GA. For $B>1$ the auxiliary space is larger than the physical one and these additional degrees of freedom, known as \textit{ghost} orbitals, increase the expressivity of the parametrization of self-energies and hybridization functions, and one recovers DMFT in the $B\rightarrow \infty$ limit.
This pole expansion representation of the self-energies and hybridization functions can be interpreted as an hybridization with fictitious auxiliary fermionic states. This way the new ghost embedding functional is given by:

\begin{align}
\mathcal{L}_\beta [\{\mathcal D_i\},\{\Lambda_i^c\},\{\mathcal R_i\},\{\Lambda_i\}]
&
=
\Omega_{\rm qp}[\mathcal R, \Lambda]
+\sum_{i=1}^{\mathcal N}\Omega_{\rm emb}^i[\mathcal D_i,\Lambda_i^c]
 -\sum_{i=1}^{\mathcal N}\Omega_{0,{\rm emb}}^i[\mathcal D_i,\Lambda_i^c;\mathcal R_i,\Lambda_i]
\label{eq:Lbeta}
\end{align}
where $\Omega_{\textrm{qp}}$, $\Omega_{\textrm{emb}}^{i}$ and $\Omega_{\textrm{0,emb}}^i$ are grand canonical potentials of systems identified by the following Hamiltonians:
\begin{align}
\hat{H}_{\text{qp}} 
    &= \sum_{ij} \sum_{a=1}^{{B} \nu_i} \sum_{b=1}^{{B} \nu_j} \hat{f}^\dagger_{i a} \, \big( \delta_{ij} [ \Lambda^i ]_{ ab} + \sum_{\alpha \beta} R_{i, a \alpha} \, t^{ij}_{\alpha\beta} \, R^\dagger_{j,\beta b}  \big) \, \hat{f}_{ j b}, \label{eq:Hqp} \\
\hat{H}^{i}_{\text{emb}} 
    &= \sum_{\alpha , \beta=1}^{\nu_i} [\epsilon_i]_{\alpha \beta } \hat{c}^\dagger_{i \alpha } \hat{c}_{i \beta} + \hat{H}^{i}_{\text{int}}[ \hat{c}^\dagger_i, \hat{c}_i ]   \nonumber\\
    &+ \sum_{ a=1}^{{B} \nu_i} \sum_{ \alpha=1 }^{\nu_i} \big( [\mathcal{D}^i]_{ a \alpha } \, \hat{b}^\dagger_{i, a} \hat{c}_{i, \alpha} + \text{h.c.} \big) + \sum_{\alpha\beta=1}^{{B} \nu_i} [ \Lambda_c^i ]_{\alpha\beta} \, \hat{f}^\dagger_\alpha \hat{b}_\beta,   \label{eq:Hemb}  \\
\hat{H}^{i}_{0,\text{emb}} 
    &= \sum_{a,b=1}^{B\nu_i} \big[\Lambda^i\big]_{ab}\, \hat{\tilde f}^\dagger_{ia} \hat{\tilde f}_{ib} +  \sum_{a,b=1}^{B \nu_i}  \Big( \big[\mathcal{D}^i (R^i)^{T}\big]_{ab}\, \hat{\tilde f}^\dagger_{ib} \hat{\tilde b}_{ia}+\text{H.c.}\Big)
    + \sum_{a,b=1}^{B\nu_i}\big[\Lambda^i_c\big]_{ab}\, \hat{\tilde b}_{ib} \hat{\tilde b}^\dagger_{ia}.  \label{eq:H0emb} 
\end{align}
where $\delta_{ij}$ is Dirac's delta.
Here, the different Hamiltonians encode the original physical fermions in different ways. In Eq.~\eqref{eq:Hemb}, the original physical fermionic fields are represented by the $\hat{c}_\alpha$ fermions. In Eqs.~(\ref{eq:Hqp},~\ref{eq:H0emb}) instead, the original physical fermionic fields are represented by linear combinations of the other fermionic fields, respectively by $\sum_a R^\dagger_{\alpha a} \hat{f}_a$ and $\sum_a R^\dagger_{\alpha a} \hat{\tilde f}_a$, more details on the role of these Hamiltonians can be found in Ref.~\cite{Giuli2026}.

\subsection{Saddle point of the ghost embedding functional}
\label{ssec:saddlepoint}

Recognizing that the ghost embedding functional can be interpreted as a sum of the grand canonical potentials drastically simplifies the saddle point conditions with respect to the four set of matrices $\{ \mathcal{R}^i , \Lambda^i, \mathcal{D}^i , \Lambda_c^i \}_{\forall i}$. These conditions are:
\begin{align}
    \frac{\partial \mathcal{L}_\beta}{\partial [\mathcal{R}_i]_{a \alpha} } = 0
        & \Rightarrow \sum_{j=1}^{\mathcal{N}} \sum_{\beta=1}^{\nu_j} \sum_{b=1}^{B \nu_j} [t^{ij}]_{\alpha \beta} [\mathcal{R}^\dagger_j ]_{\beta b} \langle \hat{f}^\dagger_{ia} \hat{f}_{j b} \rangle_{\text{qp}} 
        = \sum_{b=1}^{B \nu_i} [\mathcal{D}_i]_{b \alpha} \langle \hat{\tilde f}^\dagger_{i a} \hat{\tilde b}_{i b} \rangle_{\text{0,emb,i}} \, , \label{eq:saddle_R} \\
    \frac{\partial \mathcal{L}_\beta}{\partial [\Lambda_i]_{a b}} = 0
        &\Rightarrow \langle \hat{f}^\dagger_{ia} \hat{f}_{ib} \rangle_{\text{qp}} 
        = \langle \hat{\tilde f}^\dagger_{ia} \hat{\tilde f}_{ib} \rangle_{\text{0,emb,i}} \, , \label{eq:saddle_lambda} \\
    \frac{\partial \mathcal{L}_\beta}{\partial [\mathcal{D}_i]_{a \alpha} } = 0
        &\Rightarrow \langle \hat{c}^\dagger_{i \alpha} \hat{b}_{ib} \rangle_{\text{emb,i}} 
        = \langle \Big( \sum_{a=1}^{B\nu_i}[ \mathcal{R}_i ]_{a \alpha} \hat{\tilde f}^\dagger_{ia} \Big) \hat{\tilde b}_{ib} \rangle_{\text{0,emb,i}} \, , \label{eq:saddle_D} \\
    \frac{\partial \mathcal{L}_\beta}{\partial [\Lambda_i^c]_{a b} } = 0
        &\Rightarrow \langle \hat{b}_{ia} \hat{b}^\dagger_{ib} \rangle_{\text{emb,i}} 
        = \langle \hat{\tilde b}_{ia} \hat{\tilde b}^\dagger_{ib} \rangle_{\text{0,emb,i}}  \, . \label{eq:saddle_lambda_c}
\end{align}

The saddle point equations are now expressed as matching conditions for values of thermal one-body density matrices belonging to the different problems.
Equations~(\ref{eq:saddle_R}-\ref{eq:saddle_lambda_c}) are the self-consistency equations of the ghost embedding functional and $\langle \hat{A} \rangle_X $ is the thermal expectation value of operator $\hat{A}$ given the Hamiltonian $\hat{H}_X$ with $X$ specifying either the quasiparticle Hamiltonian or one of the embedding Hamiltonians,
more specifically $\langle \hat{A} \rangle_X \equiv \text{Tr} \left[ \hat{\rho}_X\hat{A} \right]$ with $\hat{\rho}_X \equiv \frac{e^{-\beta \hat{H}_X}}{\text{Tr} \left[ e^{-\beta \hat{H}_X} \right]}$.
When a bath size is finite, the advantage with respect to DMFT is evident: since the self-consistency is based on thermal one-body density matrices, there is no need to compute any Green's function.
This not only simplifies the self-consistency itself, since it is rooted in static expectation values instead of dynamical objects like Green's functions, but also allows the use of variational impurity solvers that have access to those thermal expectation values, such as DMRG~\cite{White1992}, tensor networks solvers, or approaches that can reduce the Hilbert space to the target manyfold of interest~\cite{Giuli2026_prxint}.

\subsection{Self-consistency cycle}
\label{ssec:selfconsistency}

\begin{figure}[h!]
    \centering
    \includegraphics[width=0.5\linewidth]{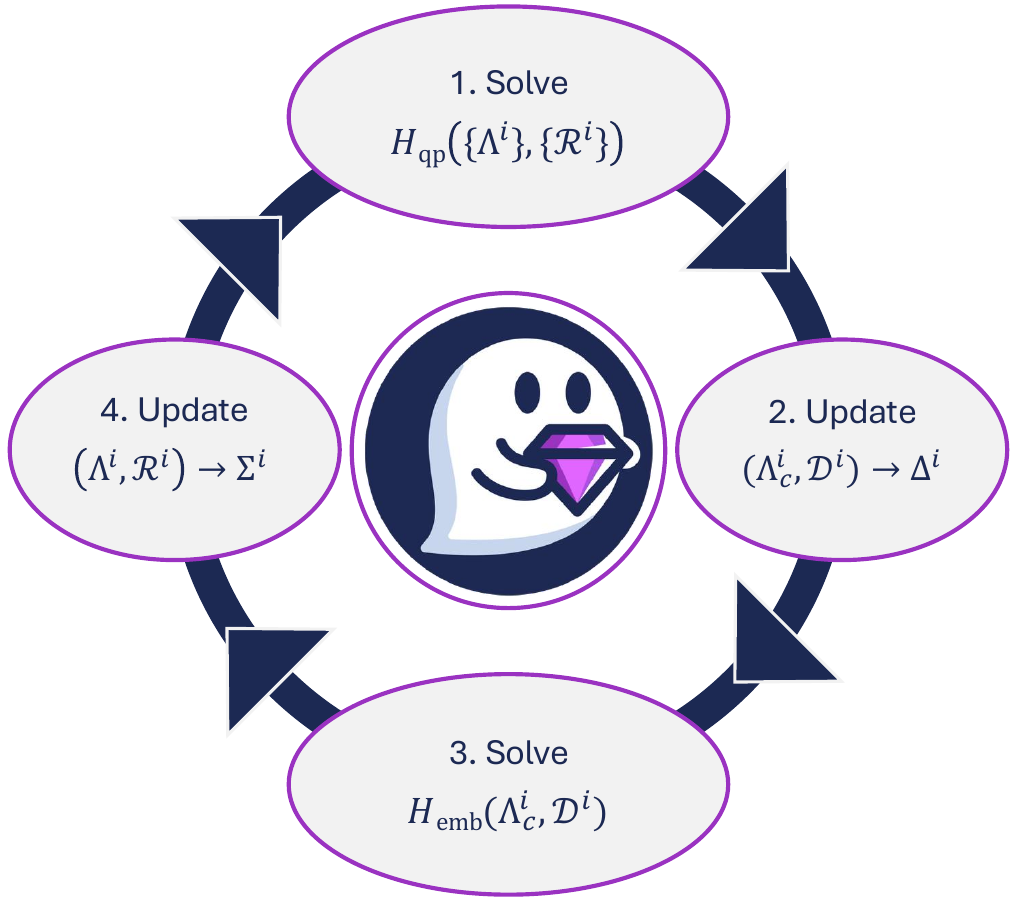}
    \caption{Schematic representation of the self-consistency cycle of the finite temperature ghost-GA implementation present in GEM with the logo of the software at the center.}
    \label{fig:selfconsistency}
\end{figure}

The self-consistency cycle of the ghost embedding functional is defined by the simultaneous fulfillment of the saddle-point Eqs.~(\ref{eq:saddle_R}-\ref{eq:saddle_lambda_c}). We implement the self-consistency by imposing some of them at each step of the self-consistency until convergence is realized. The cycle is schematically depicted in Figure~\ref{fig:selfconsistency} and can be decomposed in the following steps:
\begin{enumerate}
    \item Solve $\hat{H}_\text{qp}$ defined in Eq.~\eqref{eq:Hqp} to compute the left-hand side of Eqs.~(\ref{eq:saddle_R},~\ref{eq:saddle_lambda}).
    \label{enum:solve_qp}
    \item Update $\Delta^i[\mathcal{D}^i,\Lambda^i_c]$ to fit the right-hand side of Eqs.~(\ref{eq:saddle_R},~\ref{eq:saddle_lambda}) to the updated left side.
    \label{enum:update_hybr}
    \item Solve $\hat{H}^i_{\text{emb}}$ in Eq.~\eqref{eq:Hemb} to compute the left-hand side of Eqs.~(\ref{eq:saddle_D},~\ref{eq:saddle_lambda_c}).
    \label{enum:solve_Hemb}
    \item Update $\Sigma^i[\mathcal{R}^i,\Lambda^i]$ to fit the right-hand side of Eqs.~(\ref{eq:saddle_D},~\ref{eq:saddle_lambda_c}) to the updated left side.
    \label{enum:update_self}
\end{enumerate}
The self-consistency is reached when the matrices are converged within the tolerance set by the user and all the saddle point equations are satisfied.
At finite temperature, steps \ref{enum:update_self} and \ref{enum:update_hybr} are performed by a fitting routine for the thermal density matrices.
Details on how the thermal density-matrix fitting is performed are given in Appendix~\ref{app:denmat_fit}.
At zero temperature, as shown in Ref.~\cite{Giuli2026}, one can fix the relative gauge between the $\hat{f}$ and $\hat{b}$ fermions, and the corresponding matching conditions are realized by the ghost-GA saddle point equations. In that case, no thermal density-matrix fitting is needed.

\section{Implementation}
\label{sec:Implementation}

GEM is implemented as a Python package.
Distributed-memory parallelism is provided through a message passing interface (MPI) layer, gathered in a single \texttt{mpi.py} module that both parallel layers of the code share: the distribution of the $k$-point sums in the quasiparticle problem, provided by the \texttt{Lattice} object (See Sec.~\ref{ssec:lattice_fragment}), and the distribution of the symmetry sectors in the exact diagonalization of the embedding Hamiltonian, provided by the internal \texttt{SimpleED} solver (See Sec.~\ref{ssec:solvers}).
The module resolves the communicator, splits the work among the ranks and reduces the partial results, so that the physics routines are written once and are agnostic of the number of processes; 
the same code path is taken by a serial run, which falls back on a single-rank calculation when mpi4py~\cite{Dalcin2021mpi4py} is not available or if MPI is deactivated.
On top of this, parallelism is also inherited from the threaded \texttt{BLAS}~\cite{BLAS1990} and \texttt{LAPACK}~\cite{LAPACK1999} implementations that NumPy and SciPy link against.
The number of threads is controlled through the usual environment variables, e.g.\ \texttt{OMP\_NUM\_THREADS}.

The main software components closely follow the decomposition of the ghost-embedding functional introduced in Sec.~\ref{ssec:dynfunc}.
In particular, the lattice quasiparticle problem and the local embedding problems are represented by separate objects, called \texttt{Lattice} and \texttt{Fragment}, presented in Sec.~\ref{ssec:lattice_fragment}, while the impurity solver is exposed through an independent interface, presented in Sec.~\ref{ssec:solvers}. This separation allows the equations defining the ghost-GA saddle point to remain independent of the particular method used to solve the interacting embedding Hamiltonian.

\subsection{\texttt{Lattice} and \texttt{Fragment} objects}
\label{ssec:lattice_fragment}

The two core classes around which the software is structured are the \texttt{Lattice} and the \texttt{Fragment} classes.
The \texttt{Fragment} class contains all the information that is specific to that fragment, for example the fragment $i$ will store the matrices $\Lambda^c_i$ and $\mathcal{D}_c$ that parametrize the hybridization function $\Delta_i$, the matrices $\Lambda_i$ and $\mathcal{R}_i$ that parametrize the self-energy $\Sigma_i$, and the intra-fragment local Hamiltonian Eq.~\eqref{eq:Hloc}.
It solves the embedding problems Eqs.~(\ref{eq:Hemb},~~\ref{eq:H0emb}) and it computes local expectation values of observables.
The \texttt{Lattice} object instead contains the inter-fragment non-local kinetic terms of the Hamiltonian in Eq.~\eqref{eq:H_kin_and_loc} and solves the quasiparticle problem Eq.~\eqref{eq:Hqp}.

\paragraph{The \texttt{Fragment} class.}
At initialization time, an object belonging to this class, for a fragment labeled by $i$, will require the inputs listed in Table~\ref{tab:fragment_inputs}. We also list the corresponding physical quantity introduced in prior sections, when applicable.
The only two arguments in the Table that were not described before are: 1)~\texttt{solver}, that is the impurity solver used to solve the embedding Hamiltonian~\eqref{eq:Hemb} and is explained in more details in Sec.~\ref{ssec:solvers}, and 2)~\texttt{verbose}, a variable used to tune the verbosity of the printing from little (\texttt{verbose=0}) to very verbose (\texttt{verbose=3}).

\begin{table}[th!]
    \centering
    \begin{tabular}{c|c|c}
    \hline
    Argument & type & Corresponding quantity \\
    \hline
    \texttt{nimp} &  \texttt{int} &$\nu_i$ \\
    \texttt{nbath} &  \texttt{int} & $B\nu_i$ \\
    \texttt{eloc} &  \texttt{np.ndarray}  & $\epsilon_i$ \\
    \texttt{Utensor}   & \texttt{np.ndarray} & $U^i_{\alpha\beta\gamma\delta}$ \\
    \texttt{solver}  & See Sec.~\ref{ssec:solvers} &  \\
    \hline 
    \hline
    Optional argument & type & Corresponding quantity \\
    \hline
    \texttt{Lambda}   & \texttt{np.ndarray} & $\Lambda_i$\\
    \texttt{R}  & \texttt{np.ndarray} & $\mathcal{R}_i$\\
    \texttt{Lambda\_c}  & \texttt{np.ndarray} & $\Lambda^c_i$\\
    \texttt{D}  & \texttt{np.ndarray} & $\mathcal{D}_i$\\
    \texttt{verbose}  & \texttt{int} &  \\
    \hline
\end{tabular}
\caption{Inputs used to initialize a \texttt{Fragment} object.}
\label{tab:fragment_inputs}
\end{table}

The \texttt{Fragment} objects are then also responsible for the update of the hybridization function and self-energy parameters, e.g. steps~\ref{enum:update_hybr} and \ref{enum:update_self} in Sec.~\ref{ssec:selfconsistency} and Fig.~\ref{fig:selfconsistency}. This is done via two methods of this object, namely \texttt{update\_self\_energy} and \texttt{update\_hybridization} that automatically update the internal variables of the \texttt{Fragment}.
At finite temperature, these methods employ the thermal density-matrix fitting for which a small movement penalty, passed via the option \texttt{move\_pen} helps stabilizing the convergence. Details on the implementation of the thermal density matrix fitting are given in Appendix~\ref{app:denmat_fit}.


\paragraph{The \texttt{Lattice} class.}

At initialization time, an object belonging to the \texttt{Lattice} class will require the inputs listed in Table~\ref{tab:lattice_inputs}. We also list the corresponding physical quantity introduced in prior sections, when applicable.
The object \texttt{ek\_list} is a list of one-body Hamiltonian corresponding to the Fourier transformed superlattice hoppings $t^K$, as defined in Eq.~\eqref{eq:H_kin_and_loc_supercell_K}.
One can optionally pass an array \texttt{wk\_list} which contains the multiplicity of each element of the \texttt{ek\_list} to enable density of state calculations or restricting to the irreducible part of the Brillouin zone.

The two remaining optional arguments control the parallel layer: \texttt{use\_mpi} switches the distribution of the $k$-points off, and \texttt{comm} allows one to pass an explicit communicator, e.g.\ a sub-communicator nesting the $k$-point splitting inside another level of parallelism. Left to their defaults, the code uses \texttt{MPI.COMM\_WORLD} when mpi4py is available and a single-rank stand-in otherwise, so that a serial run needs no change.
The sums over $k$ are distributed over the MPI ranks, each rank owning a contiguous block of $k$-points of equal size up to one, and closed by an \emph{allreduce} so that all ranks share the same result.

\begin{table}[th!]
    \centering
    \begin{tabular}{c|c|c}
    \hline
    Argument  & type & Corresponding quantity \\
    \hline
    \texttt{ek\_list} & \texttt{np.ndarray} & $\{ t^K \}_K$ \\
    \hline
    \hline
    Optional argument  & type & Corresponding quantity \\
    \hline
    \texttt{wk\_list}  & \texttt{np.ndarray} &  \\
    \texttt{verbose}   & \texttt{int} & \\
    \texttt{use\_mpi}  & \texttt{bool} & \\
    \texttt{comm}      & \texttt{MPI.Comm} & \\
    \hline
\end{tabular}
\caption{Inputs used to initialize a \texttt{Lattice} object.}
\label{tab:lattice_inputs}
\end{table}

\subsection{Solvers: the \texttt{gemSolver} class, \texttt{SimpleED} and generic template}
\label{ssec:solvers}

Solving the interacting embedding Hamiltonian of Eq.~\eqref{eq:Hemb} is the time bottleneck of the method. GEM therefore keeps the solution of the embedding problem behind a small, explicit interface, so that the self-consistency machinery is independent of the strategy used to solve it. A solver has to be passed to each \texttt{Fragment} object when the latter is initialized, and the \texttt{Fragment} verifies that it is an instance of \texttt{gemSolver}.

\paragraph{The \texttt{gemSolver} class.}
Every solver inherits from \texttt{gemSolver}, whose constructor stores the two
attributes shared by all solvers: 1)~\texttt{type}, a label used in the printouts that specify the name of the solver,
and 2)~\texttt{solver\_params}, a dictionary of solver-specific parameters. A subclass must implement four methods, which are the only pieces that the rest of the code requires:
\begin{itemize}
  \item \texttt{build\_Hemb(D, eloc, Lambdac, Utensor)}, which constructs the embedding Hamiltonian from the four tensors that define it;
  \item \texttt{solve\_Hemb(T, verbose)}, which solves the embedding Hamiltonian defined before at a temperature \texttt{T}. This method also sets the attributes \texttt{gs\_ene}, the ground-state energy, and \texttt{Zpart}, the partition function normalized so that it equals one at zero temperature;
  \item \texttt{calc\_density\_matrix()}, which calculates and returns the one-body density matrix $\langle c^\dagger_i c_j\rangle$ of the impurity and bath degrees of freedom;
  \item \texttt{compute\_E2loc()}, which calculates and returns the local two-body energy.
\end{itemize}

Two further methods, \texttt{compute\_E1loc(nimp)} and
\texttt{calc\_double\_occ()}, are optional and only needed by the routines that evaluate the corresponding observables (the local one-body energy and the double occupation). Everything else is left to the solver.
In particular, the parameters controlling the algorithm are never passed by the \texttt{Fragment}: they are fixed once, when the solver is constructed, and read from \texttt{solver\_params}, which makes the interface independent of the solver in use.

\paragraph{The \texttt{SimpleED} solver.}
GEM ships with an internal exact-diagonalization solver: \texttt{SimpleED}, which implements \textsc{LAPACK} based full diagonalization and ARPACK, both via SciPy~\cite{2020SciPy-NMeth}.
It provides the possibility of using symmetry sectors, with the total number of electrons and the total magnetization as good quantum numbers, activated at initialization by passing respectively \texttt{use\_Ntot=True} and \texttt{use\_Sz=True}.
The user can also fix the symmetry sector by passing the desired list of eigenvalues through the options \texttt{N\_sector} and \texttt{Sz\_sector}; leaving them to \texttt{None} includes all sectors, which is the correct choice at finite temperature.
The \texttt{solver\_params} dictionary collects the parameters of the diagonalization and of the sampling of the thermal states: the number of eigenpairs \texttt{solver\_params['num\_eig']}, which defaults to the ground state alone at $T=0$ and to the full spectrum at $T>0$; the dimension \texttt{solver\_params['dense\_cutoff']} above which a sector is diagonalized with ARPACK rather than fully; the ARPACK options \texttt{solver\_params['which']} and \texttt{solver\_params['tol']}, controlling respectively the type of method used and the tolerance; the smallest Boltzmann weight \texttt{solver\_params['bw\_cutoff']} retained in the partition function.

\texttt{SimpleED} also implements a matrix-free version of ARPACK meant for calculations with large Hilbert spaces that do not allow storing the full sparse matrices. The matrix-free version can be activated using \texttt{solver\_params['matrix\_free'] = True}.  
The symmetry sectors are independent of one another, and they are distributed over the ranks of the MPI communicator, passed through the \texttt{comm} option and defaulting to \texttt{MPI.COMM\_WORLD} (the distribution is switched off by \texttt{solver\_params['use\_mpi'] = False}).
The sectors differ widely in size, so the assignment is not uniform as for the $k$-points, but a greedy longest-processing-time-first packing: the sectors are sorted by decreasing estimated cost $\text{dim}^{p}$ and each is handed to the least loaded rank, with the exponent $p$ set by \texttt{solver\_params['mpi\_weight\_exp']}, defaulting to $3$ for dense diagonalization and $1$ in matrix-free mode, where the cost is a number of matrix-vector products.
Each rank then builds the basis and the operators only for the sectors it owns, so that memory scales down with the number of ranks as well, while the partition function, the ground-state energy and the thermal expectation values are obtained by reduction over the ranks and are identical everywhere.
The routines involving such reductions are therefore collective and are called by every rank.

\paragraph{Adding a new solver.}
The file \texttt{solver\_template.py} contains a commented skeleton that can be copied to interface an external solver: it shows how to inherit from \texttt{gemSolver}, marks which arguments are passed by the \texttt{Fragment} and which are solver-specific, and documents the quantities each method has to return.
When a solver other than \texttt{SimpleED} is used, the appropriate references should be cited alongside the GEM one.

\subsection{Testing suit}
\label{ssec:testing}

The software comes with unit testing and smoke testing. 
As shown in Section~\ref{sec:Installation}, the testing suit is built by default at installation time, unless explicitly turned off. From the building directory, one can execute the full suit of tests by running \texttt{make test}

\section{Examples}
\label{sec:examples}

In this section, we present the functionalities of GEM with a set of paradigmatic calculations for strongly correlated systems.
The Python scripts to run these examples can be found in the \texttt{gem/examples} folder. The second example is the only one using an external library, \textsc{triqs}~\cite{Parcollet2015}, for the construction of the non interacting Hamiltonian. Nonetheless, it is not necessary and it is just meant to show the possibility of connecting our software with external established packages such as \textsc{triqs}.
The remaining examples are all dependent on only GEM.
The example in Sec.~\ref{ssec:E1-Bethe-Tfinite} presents a finite temperature calculation for the single orbital Hubbard model where the total energy, entropy and double occupancy are evaluated. The example in Sec.~\ref{ssec:E2-Square-Tfinite} presents a study of the magnetic phase diagram of the square lattice Hubbard model with two inequivalent fragments for the inequivalent sublattice sites. This example determines the critical temperatures and compares the phase diagram with DMFT results in literature. Finally, the example in Sec.~\ref{ssec:E3-Bethe-multiorb-Tzero} presents a three-orbitals calculation for a Hubbard-Kanamori interaction on the Bethe lattice, to identify Hund's physics at fractional filling.

\subsection{Bethe lattice: thermodynamics}
\label{ssec:E1-Bethe-Tfinite}

In this example, we show how to perform finite temperature calculations for the single-orbital Hubbard model on the Bethe lattice.
We choose the half-bandwidth ($D$) of the semi-circular density of state as the unit of energy and we show how to compute the total energy per site and the entropy per site at half-filling for $U/D=2.0$. The script performing this calculation can be found in \texttt{gem/examples/E1/Bethe\_1orb.py}.
We start from importing the relevant packages:
\begin{lstlisting}[style=mypython,numbers=none]
import numpy as np
from gem.fragment import Fragment
from gem.lattice import Lattice
from gem.solvers.simple_ed import SimpleED
\end{lstlisting}
Next, we define the parameters of the model and the self-consistency. The model has one band and two spins giving \texttt{nimp = 2}, with three ghosts per physical orbital $B=3$.
We choose $U/D=2.0$ and, since the Bethe lattice density of states is particle-hole symmetric, we enforce half-filling by fixing $\mu=U/2$.
We choose a set of temperatures between $T/D=0.001$ and $T/D=1.0$ on a log grid, in addition to $T=0$ as the first temperature: 
\begin{lstlisting}[style=mypython,numbers=none]
# Parameters that determine the physical dimentions 
# 1 orbital with 2 spins, B=3 bath per orbital, total 8
B, nimp = 3, 2
nbath = nimp*B
ntot = nimp+nbath

# Physical Parameters
U, mu = 2.0, 1.0
Tlist = np.hstack((np.array([0]), np.logspace(np.log10(1e-3), np.log10(1), 31)))

# Self-consistency Parameters
itmax, tol, Tsmearing = 100, 1e-5, 1e-3
\end{lstlisting}
The last line is additional parameters: we perform a maximum of \texttt{itmax} iterations, unless we reach the convergence criterion set by \texttt{tol}.
Moreover, we use a small smearing for the temperature \texttt{Tsmearing} so that the Fermi function of the \textit{quasiparticle} problem does not have a jump at zero temperature.

We then define the kinetic part of the Hamiltonian and initialize a \texttt{Lattice} object containing this information:
\begin{lstlisting}[style=mypython,numbers=none]
# Non-interacting density of states and lattice object for the Bethe lattice
e_list = np.linspace(-1, 1, 5001)
wks = np.sqrt(1 - e_list**2)
wks /= np.sum(wks)
eks = e_list[:, None, None] * np.eye(2, dtype=np.complex128)

lattice = Lattice(eks, wk_list=wks)
\end{lstlisting}
and the local Hamiltonian together with a \texttt{SimpleED} solver and a \texttt{Fragment} object:
\begin{lstlisting}[style=mypython,numbers=none]
# Local Hamiltonian
eloc = np.zeros((nimp,nimp))
# Interaction tensor of the embedded space
Utensor = np.zeros((nimp, nimp, nimp, nimp))
Utensor[0,0,1,1], Utensor[1,1,0,0] = U, U

# SimpleED solver initialization
edsolver = SimpleED(ntot, use_Ntot=True, use_Sz=True, N_sector=None, Sz_sector=None, 
                    dtype=np.float64)

# Fragment initialization
Lambda0 = None; R0 = None
fragment = Fragment(nimp, nbath, eloc, Utensor, edsolver, Lambda=Lambda0, R=R0, 
                    verbose=2)
\end{lstlisting}
We define a function to check for convergence of $\mathcal R$ and $\Lambda$ matrices. Due to the gauge freedom of these matrices, we check convergence for gauge invariant quantities:
\begin{lstlisting}[style=mypython,numbers=none]
def check_convergence(R_new, L_new, R_old, L_old):
    # Only 1 spin and gauge invariant difference
    L_eval_new, UL_new = np.linalg.eigh(L_new[::2,::2])
    L_eval_old, UL_old = np.linalg.eigh(L_old[::2,::2])
    R_eval_old, R_eval_new = UL_old @ R_old[::2,::2], UL_new @ R_new[::2,::2]
    diff_R = np.abs(np.abs(R_eval_old) - np.abs(R_eval_new)).max()
    diff_Lambda = np.abs(L_eval_new - L_eval_old).max()
    diff = max(diff_R, diff_Lambda)
    return diff

\end{lstlisting}
And finally, we setup a self-consistency cycle storing the total energy, entropy and double occupancies:
\begin{lstlisting}[style=mypython,numbers=none]
Elist, Slist, docclist = [], [], []

for T in Tlist:
    print('--------------------------------------')
    print(f'GEM loop started with U={U} and T={T}')

    # Self consistency loop
    for it in range(itmax):
        print(f'----- ghost-RISB iteration {it} / {itmax} -----')
        lattice.solve_qp([fragment], T=T, Tsmearing=Tsmearing) # Step 1: Solve H_qp
        fragment.update_hybridization(T=T, use_Sz=True) # Step 2: Update Delta
        fragment.solve_impurity(mu, T=T) # Step 3: Solve H_emb
        Lambda_old, R_old = fragment.Lambda.copy(), fragment.R.copy() # Save old Sigma
        fragment.update_self_energy(T=T, use_Sz=True) # Step 4: Update Sigma
        Lambda_new, R_new = fragment.Lambda, fragment.R # Save new Sigma

        diff = check_convergence(R_new, Lambda_new, R_old, Lambda_old)
        print(f'iteration: {it}  diff={diff}')

        if (diff < tol and it > 2) or it == itmax - 1:
            print(f'----- Exiting loop with diff={diff} after '
                  f'{it} iterations (max={itmax}) -----'); break

    docc = fragment.E2loc/U
    docclist.append(docc)
    ekin = lattice.compute_ekin([fragment], T=T, Tsmearing=Tsmearing)
    eimp = fragment.compute_energy()
    etot = (eimp+ekin).real
    # Free-energy functional
    L = lattice.compute_functional([fragment], T=T, Tsmearing=Tsmearing).real
    S = (etot-L)/T if T > 0 else 0.0

    Elist.append(etot); Slist.append(S)

    print('--------------------------------------')
    print(f'GEM loop ended with U={U} and T={T}')
    print(f'returning docc={docc}, Etot={etot} and S={S}')
    print('--------------------------------------')

np.savetxt('Tlist.dat', Tlist); np.savetxt('Elist.dat', Elist)
np.savetxt('Slist.dat', Slist); np.savetxt('docclist.dat', docclist)
\end{lstlisting}

\begin{figure}[ht!]
    \centering
    \includegraphics[width=\linewidth]{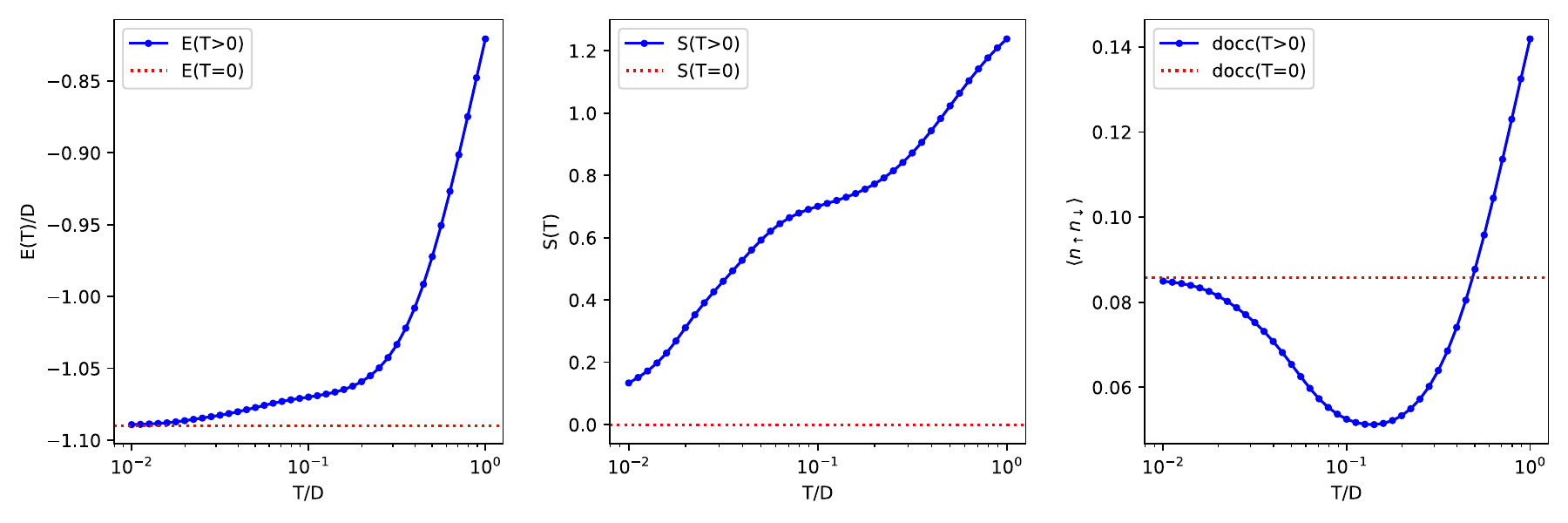}
    \caption{Temperature evolution of the total energy per site $E$ (left), entropy per site $S$ (center) and double occupancy (right) for the Hubbard model on the Bethe lattice with $U/D=2.0$ using $B=3$.}
    \label{fig:E1_all}
\end{figure}

In Fig.~\ref{fig:E1_all}, we report the behavior of the total energy per site ($E$), entropy per site ($S$) and double occupancies ($\langle n_{\uparrow}n_{\downarrow}\rangle$) as a function of temperature ($T$). The $T=0$ values (red dotted line) are compatible with the finite temperature ones (blue line). This is in agreement with Ref.~\cite{Giuli2026} where a comparison with DMFT was also given.

\subsection{Square lattice: magnetic phase diagram}
\label{ssec:E2-Square-Tfinite}

In this example, we show how to perform finite-temperature calculations for computing the magnetic phase diagram of the single-orbital Hubbard model on the square lattice at half-filling. We use two inequivalent fragments as an example of a multi-fragment calculation.
We choose the half-bandwidth ($D=4t$) as the unit of energy, therefore $t=0.25$, and compute the antiferromagnetic order parameter along the $z$-axis, that is $\langle n_{\uparrow} - n_{\downarrow} \rangle$, as a function of temperature. We then extract the critical temperature ($T_c$) and compare with DMFT data obtained from Ref.~\cite{Kunes2011}.
We make use of \textsc{triqs}~\cite{Parcollet2015} to generate the dispersion \texttt{ek\_list} from a tight-binding object, which is then passed to the \texttt{Lattice} object. The script performing this calculation can be found in \texttt{gem/examples/E2/Square\_1orb\_2frag\_phase.py}.

We start again from importing the relevant packages:
\begin{lstlisting}[style=mypython,numbers=none]
import numpy as np
from gem.fragment import Fragment
from gem.lattice import Lattice
from gem.solvers.simple_ed import SimpleED
import h5py
\end{lstlisting}
We then define the parameters of the model and the self-consistency. We present a script performing a calculation for a single $U$. We impose half-filling by setting the chemical potential to be $\mu = U/2 $. The temperatures are linearly spaced between $T/D=0$ and $T/D=0.1$. We set the hopping to $t=0.25$ and include a small magnetic field \texttt{bfield} that helps stabilizing the antiferromagnetic solution at low temperatures for the first few iterations. Moreover, we use a small linear mixing \texttt{mix} for $\Lambda$ and $\mathcal{R}$ and,  as shown in Appendix~\ref{app:denmat_fit}, we use a small movement penalty \texttt{move\_pen} for the thermal density matrix fitting that helps stabilizing convergence.
\begin{lstlisting}[style=mypython,numbers=none]
# Parameters that determine the physical dimensions
# 1 orbital with 2 spins, B=3 bath per orbital, total 8
nimp, B = 2, 3
nbath = nimp * B
ntot = nimp + nbath

# Physical Parameters
U, mu, t, bfield = 0.8, 0.4, 0.25, 1e-2
T_list = np.linspace(0.0, 0.10, 51)

# Self-consistency Parameters
itmax, tol, mix, move_pen, Tsmearing = 200, 1e-3, 0.05, 1e-8, 1e-3
# Output file
h5_file = f'data_Square_1orb_2frag_B{B}_phase.h5'
\end{lstlisting}

We then make use the \texttt{TBLattice} object from \textsc{triqs} to generate the dispersion and initialize the \texttt{Lattice} object:
\begin{lstlisting}[style=mypython,numbers=none]
from triqs.lattice.tight_binding import TBLattice

# Two-site (Neel) unit cell of the square lattice: A at (0,0), B at (1,0),
# primitive vectors along the diagonals so that the two sublattices alternate.
a1, a2 = np.array([1.0,  1.0, 0.0]), np.array([1.0, -1.0, 0.0])
H_t = TBLattice(
    units=[a1, a2],
    hoppings={( 0,  0): [[0, -t], [-t, 0]],
              (-1, -1): [[0, -t], [ 0, 0]], ( 1,  1): [[0,  0], [-t, 0]],
              ( 0, -1): [[0, -t], [ 0, 0]], ( 0,  1): [[0,  0], [-t, 0]],
              (-1,  0): [[0, -t], [ 0, 0]], ( 1,  0): [[0,  0], [-t, 0]]},
    orbital_positions=[(0, 0, 0), (1, 0, 0)], orbital_names=['A', 'B'],
)

# Build the k-grid
Nk = 200
kmesh = H_t.get_kmesh((Nk, Nk, 1))
kpts_cart = np.array(list(kmesh.values()))
recip = np.array(kmesh.bz.units)              # rows: reciprocal lattice vectors
kpts = kpts_cart @ np.linalg.inv(recip)

# List of H(k) matrices on the square-lattice BZ, one spin block per sublattice
eks = np.array([np.kron(H_t.fourier(k), np.eye(2)) for k in kpts])
lattice = Lattice(eks)
\end{lstlisting}
We then define the local Hamiltonian and initialize solvers and fragment objects for each sublattice:
\begin{lstlisting}[style=mypython,numbers=none]
# Local Hamiltonian
eloc = np.zeros( (nimp,nimp) )
# Interaction tensor of the embedded space
Utensor = np.zeros((nimp, nimp, nimp, nimp))
Utensor[0, 0, 1, 1], Utensor[1, 1, 0, 0] = U, U

# Let the fragment decide a starting value
Lambda_A, R_A = None, None;   Lambda_B, R_B = None, None
Lambda_c_A, D_A = None, None; Lambda_c_B, D_B = None, None

#Initalize solvers and fragments for the two sublattices
edsolverA = SimpleED(ntot, use_Ntot=True, use_Sz=True, dtype=np.complex128)
edsolverB = SimpleED(ntot, use_Ntot=True, use_Sz=True, dtype=np.complex128)
fragmentA = Fragment(nimp, nbath, eloc, Utensor, edsolverA, verbose=0,
                     Lambda=Lambda_A, R=R_A, Lambda_c=Lambda_c_A, D=D_A)
fragmentB = Fragment(nimp, nbath, eloc, Utensor, edsolverB, verbose=0,
                     Lambda=Lambda_B, R=R_B, Lambda_c=Lambda_c_B, D=D_B)

\end{lstlisting}
We then create the output file and an array to store the magnetizations and other quantities:
\begin{lstlisting}[style=mypython,numbers=none]
# create the file if not present; each U group is appended as it completes
with h5py.File(h5_file, 'a') as _:  # 'a' append, 'w' write and delete previous
    pass

# mag_grid[:,iT] = [ m_A, m_B ] order parameter on sublattice A and B
mag_grid = np.zeros( (2,len(T_list)) )
magA = 0.0; magB = 0.0
\end{lstlisting}
Finally, we write the self-consistency cycle checking the convergence on both the $\Lambda$ and $\mathcal R$ matrices,
as well as on the magnetization on each sublattice:
\begin{lstlisting}[style=mypython,numbers=none]
for iT, T in enumerate(T_list):
    print('--------------------------------------------------------')
    print(f'GEM loop started with U={U} and T={T}')

    # Self consistency loop
    for it in range(itmax):
        print(f"----- ghost-RISB iteration {it} / {itmax} -----")

        lattice.solve_qp([fragmentA, fragmentB], T=T, Tsmearing=Tsmearing)

        fragmentA.update_hybridization(T=T, use_Sz=True, move_pen=move_pen)
        fragmentB.update_hybridization(T=T, use_Sz=True, move_pen=move_pen)

        # seed AFM only at lowest T for the first few iterations
        if iT == 0 and it < 3:
            fragmentA.eloc = eloc + bfield * np.diag([-1, 1])
            fragmentB.eloc = eloc - bfield * np.diag([-1, 1])
        else:
            fragmentA.eloc = eloc.copy()
            fragmentB.eloc = eloc.copy()

        fragmentA.solve_impurity(mu, T=T)
        fragmentB.solve_impurity(mu, T=T)

        Lambda_old_A = fragmentA.Lambda.copy(); R_old_A = fragmentA.R.copy()
        Lambda_old_B = fragmentB.Lambda.copy(); R_old_B = fragmentB.R.copy()

        fragmentA.update_self_energy(T=T, use_Sz=True, move_pen=move_pen)
        fragmentB.update_self_energy(T=T, use_Sz=True, move_pen=move_pen)

        diff_LR = max(
            np.abs(fragmentA.Lambda - Lambda_old_A).max(),
            np.abs(np.abs(fragmentA.R) - np.abs(R_old_A)).max(),
            np.abs(fragmentB.Lambda - Lambda_old_B).max(),
            np.abs(np.abs(fragmentB.R) - np.abs(R_old_B)).max(),
        )

        # Mixing of Lambda and R
        fragmentA.Lambda = (1 - mix) * fragmentA.Lambda + mix * Lambda_old_A
        fragmentA.R = (1 - mix) * fragmentA.R + mix * R_old_A
        fragmentB.Lambda = (1 - mix) * fragmentB.Lambda + mix * Lambda_old_B
        fragmentB.R = (1 - mix) * fragmentB.R + mix * R_old_B

        magA_old, magB_old = magA, magB
        magA = (fragmentA.denMat[0, 0].real - fragmentA.denMat[1, 1].real)
        magB = (fragmentB.denMat[0, 0].real - fragmentB.denMat[1, 1].real)

        diff_mA = abs(magA - magA_old)
        diff_mB = abs(magB - magB_old)
        diff = max(diff_LR, 10 * diff_mA, 10 * diff_mB)

        print(f'iteration: {it}  diff={diff:.2e} '
                f'diff_mA={diff_mA:.4f} diff_mB={diff_mB:.4f} '
                f'magA={magA:.4f} magB={magB:.4f} '
                f'nfillA={fragmentA.nfill:.4f} nfillB={fragmentB.nfill:.4f} ')

        if (diff < tol and it > 1) or it == itmax - 1:
            print(f"--- Exiting after {it}/{itmax} - diff={diff:.2e}  ---")
            break

    magA = (fragmentA.denMat[0,0].real - fragmentA.denMat[1,1].real)
    magB = (fragmentB.denMat[0,0].real - fragmentB.denMat[1,1].real)
    mag_grid[0,iT] = magA; mag_grid[1,iT] = magB


    print('--------------------------------------------------------')
    print(f'GEM loop ended with U={U} and T={T}')
    print(f'returning m_A={mag_grid[0,iT]} and m_B={mag_grid[1,iT]}')
    print('--------------------------------------------------------')
\end{lstlisting}
After convergence we save the data in the output file:
\begin{lstlisting}[style=mypython,numbers=none]
# save to output file
with h5py.File(h5_file, 'a') as h5f:
    grp_name = f'U{U:.2f}_B{B}'
    if grp_name in h5f:
        del h5f[grp_name]
    grp = h5f.create_group(grp_name)
    grp.create_dataset('T_list',   data=T_list)
    grp.create_dataset('mag_grid', data=mag_grid)
print(f'Written U={U:.2f} to {h5_file}')
\end{lstlisting}

\begin{figure}[h!]
    \centering
    \includegraphics[width=\linewidth]{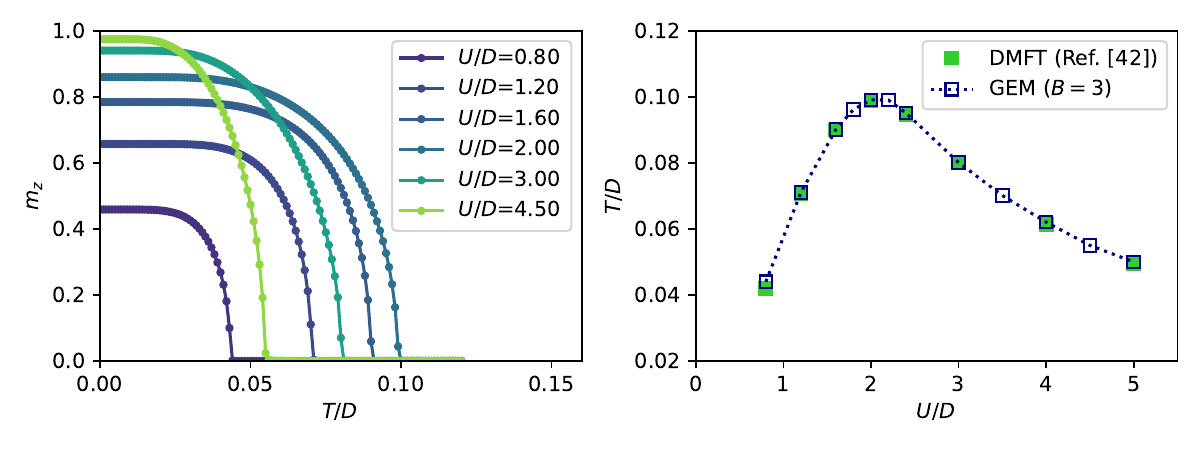}
    \caption{Antiferromagnetic order parameter (left) as a function of temperature for different values of $U/D$ obtained with GEM using $B=3$, and temperature versus interaction antiferromagnetic phase boundaries (right) for the square lattice Hubbard model compared with DMFT from Ref.~\cite{Kunes2011}.  }
    \label{fig:E2_all}
\end{figure}

In Fig.~\ref{fig:E2_all} (left), we plot the antiferromagnetic order parameter defined as the magnetization per site ($ m_z = \langle n_{i A\uparrow  } - n_{i A\downarrow  } \rangle =\langle n_{i B\downarrow  } - n_{i B\uparrow  } \rangle$).

We use $B=3$ and see that it correctly captures the BCS-BEC crossover with a maximum critical temperature reached for $U/D \approx 2$.
In Figure~\ref{fig:E2_all}(right), we compare the phase boundaries obtained by GEM with the DMFT phase boundaries obtained from Ref.~\cite{Kunes2011}.

\subsection{Multi-orbital Bethe lattice: Hund's physics}
\label{ssec:E3-Bethe-multiorb-Tzero}

In this example, we show how to perform zero temperature calculations for a multi-orbital system with Hubbard-Kanamori rotationally-invariant interaction~\cite{Georges2013}. We focus on 3-orbitals on the Bethe lattice and show how the quasiparticle weight evolve with interaction and the increase of the critical $U$ for values of $J/U$ close to $1/3$.
We choose the half-bandwidth ($D$) of the semi-circular density of state as the unit of energy and we compare with GA and ghost-GA calculations found in literature, in particular Refs.~\cite{Lanata2013,MejutoZaera2023}.
The script performing this calculation can be found in \texttt{gem/examples/E3/Bethe\_3o.py}.

We start from importing the relevant packages:
\begin{lstlisting}[style=mypython,numbers=none]
import numpy as np
from gem.fragment import Fragment
from gem.lattice import Lattice
from gem.solvers.simple_ed import SimpleED
from gem.utilities import U_matrix_kanamori
\end{lstlisting}
Note that we imported the \texttt{U\_matrix\_kanamori} function implemented in the GEM utilities module.
Next, we define the parameters of the model and of self-consistency:
\begin{lstlisting}[style=mypython,numbers=none]
n_orb, B = 3, 3 # B will be 1 or 3 in this example
nimp = 2 * n_orb
nbath = nimp * B
ntot = nimp + nbath

# Physical Parameters
U_list, JoverU_list = np.linspace(0.1,10.0,100), np.array([0.00, 0.10, 0.20, 0.30])
T, n_target = 0, n_orb-1

# Self-consistency Parameters
itmax, tol, mix, ntol, Tsmearing = 200, 1e-4, 0.2, 1e-4, 1e-3
\end{lstlisting}
In this calculation, \texttt{tol} is the tolerance on the change of $\Lambda$ and $R$, that serves as the convergence criterion for the self-consistency. The variable \texttt{ntol} is the tolerance on the impurity filling.

We then define the kinetic part of the Hamiltonian and initialize a \texttt{Lattice} object containing this information:
\begin{lstlisting}[style=mypython,numbers=none]
# Non-interacting density of states and lattice object
e_list = np.linspace(-1, 1, 1001)
wks = np.sqrt(1 - e_list**2)
wks /= np.sum(wks)
eks = e_list[:, None, None] * np.eye(nimp, dtype=np.complex128)
lattice = Lattice(eks, wk_list=wks)
\end{lstlisting}
and a function to check the convergence:
\begin{lstlisting}[style=mypython,numbers=none]
def check_convergence(R_new, L_new, R_old, L_old):
    # Only 1 spin and gauge invariant difference
    L_eval_new, UL_new = np.linalg.eigh(L_new[::2*n_orb, ::2*n_orb])
    L_eval_old, UL_old = np.linalg.eigh(L_old[::2*n_orb, ::2*n_orb])
    diff_R = ( np.abs(np.abs(UL_old @ R_old[::2*n_orb, ::2*n_orb])
         - np.abs(UL_new @ R_new[::2*n_orb, ::2*n_orb])) ).max() 
    diff_Lambda = np.abs(L_eval_new - L_eval_old).max()
    diff = max(diff_R, diff_Lambda)
    return diff
\end{lstlisting}
We initialize arrays to store the information for the future plots and then built the cycle similarly to what was shown in the previous examples:
\begin{lstlisting}[style=mypython,numbers=none]
Zgrid = np.zeros((len(JoverU_list), len(U_list)))
ngrid = np.zeros((len(JoverU_list), len(U_list)))

for iJ, JoverU in enumerate(JoverU_list):
    # warm-start: carry the converged Lambda/R and mu from one U to the next
    Lambda0, R0, mu = None, None, 0
    for iU, U in enumerate(U_list):
        J = JoverU * U
        print('--------------------------------------------------------')
        print(f'GEM loop started with U={U:.2f} and J={J:.2f} (J/U={JoverU:.2f})')

        # Local Hamiltonian: the level position is set by mu, fitted to n_target
        eloc = np.zeros((nimp, nimp))
        # Interaction tensor of the embedded space
        Utensor = U_matrix_kanamori(n_orb, U, J) 

        # Fixed sectors for T=0 calculations
        edsolver = SimpleED(ntot, use_Ntot=True, use_Sz=True,
                            N_sector=ntot//2, Sz_sector=0, dtype=np.float64)

        # Fragment initialization
        fragment = Fragment(nimp, nbath, eloc, Utensor, edsolver,
                            Lambda=Lambda0, R=R0, verbose=0)

        # Self consistency loop
        for it in range(itmax):
            print(f"----- ghost-RISB iteration {it} / {itmax} -----")
            lattice.solve_qp([fragment], T=T, Tsmearing=Tsmearing)
            fragment.update_hybridization(T=T, use_Sz=True)
            fragment.solve_impurity(mu, T=T)

            #Fit density to n_target
            if abs(fragment.nfill.real - n_target) > ntol:
                mu = lattice.fit_mu(n_target, [fragment], T=T, mu_old=mu,
                                    mode='imp', ntol=1e-5)
                fragment.solve_impurity(mu, T=T)
            nfill=fragment.nfill.real

            Lambda_old = fragment.Lambda.copy(); R_old = fragment.R.copy()
            fragment.update_self_energy(T=T, use_Sz=True)
            # stay in the paramagnetic, orbitally degenerate solution
            fragment.impose_orbital_symmetry()
            fragment.impose_spin_SU2_symmetry()
            Lambda_new = fragment.Lambda; R_new = fragment.R

            diff = check_convergence(R_new, Lambda_new, R_old, Lambda_old)
            print(f"it:{it}  diff={diff:.3e}  mu={mu:.4f}  nfill={nfill:.4f}")

            fragment.Lambda = (1 - mix) * Lambda_new + mix * Lambda_old
            fragment.R = (1 - mix) * R_new + mix * R_old

            if (diff < tol and it > 2) or it == itmax - 1:
                print(f"-- Exiting with diff={diff:.3e} after {it} iterations")
                break

        Z = fragment.compute_Z()
        Zgrid[iJ, iU], ngrid[iJ, iU] = Z.real[0, 0], fragment.nfill.real

        # warm start for the next U
        Lambda0, R0 = fragment.Lambda.copy(), fragment.R.copy()

        print('--------------------------------------------------------')
        print(f'GEM loop ended with U={U} and J={J}')
        print(f'returning Z={np.diag(Z.real)}, mu={mu} and n={fragment.nfill.real}')
        print('--------------------------------------------------------')

np.savetxt('hund_Ulist.dat', U_list); np.savetxt('hund_JoverU.dat', JoverU_list)
np.savetxt('hund_Zgrid.dat', Zgrid);  np.savetxt('hund_ngrid.dat', ngrid)
\end{lstlisting}

\begin{figure}[h!]
    \centering
    \includegraphics[width=\linewidth]{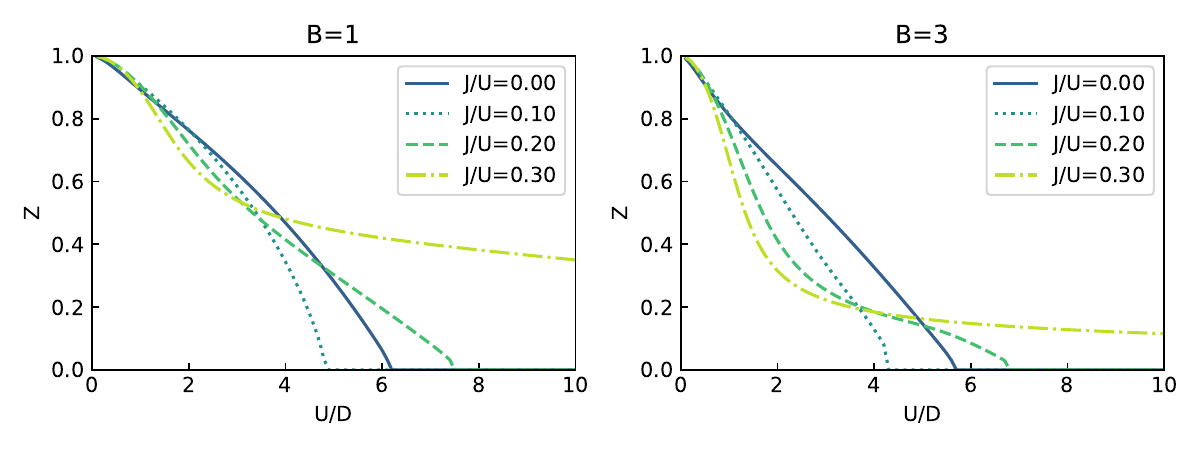}
    \caption{Quasiparticle weight $Z$ for a three-orbital model with local Hubbard-Kanamori interaction on the Bethe lattice at filling $n$=2 as a function of $U/D$ and for $J/D=0.0, 0.1, 0.2$ and $0.3$ using GEM, for $B=1$ (left panel) and $B=3$ (right panel).}
    \label{fig:E3_B1}
\end{figure}

In Fig.~\ref{fig:E3_B1}, we show the results of the calculations for $J/U = 0.0, 0.1, 0.2$ and $0.3$, where we plot the quasiparticle weight $Z$ as a function of $U/D$ for $B=1$ (left panel) and $B=3$ (right panel). The results can be compared with Ref.~\cite{Lanata2013} for $B=1$ and with Ref.~\cite{MejutoZaera2023} for $B=3$. When Hund's physics becomes important around $U/J \approx1/3$ in this system, $B=3$ shows a much larger renormalization of the quasiparticle weight and is in much better agreement with DMFT, as shown in Ref.~\cite{MejutoZaera2023}.

\section{Outlook}
\label{sec:outlook}

The modular structure of GEM provides a natural framework for extending its capabilities and interfacing it with other computational methods and software packages, and several directions for future development are being explored.

One direction concerns the integration of additional impurity solvers. In particular, interfaces to ground-state solvers based on density-matrix renormalization group (DMRG) and matrix product states (MPS) methods, coupled-cluster approaches, and other approximate ground-state solvers~\cite{Giuli2026_prxint} could considerably broaden the range of systems and regimes accessible within GEM.

Another extension is the inclusion of retarded interactions, which would allow GEM to address a wider class of effective low-energy models with frequency-dependent interactions that have been shown to be a promising route for this method~\cite{Giuli2026eph}.

Finally, interfacing GEM with \textit{ab initio} electronic-structure codes will provide a route towards material-specific calculations. At the moment, an interface with \textsc{triqs/ModEST}~\cite{modest26} is under development, enabling DFT+ghost-GA calculations, which have already proven to be a cheap and fast alternative to DFT+DMFT~\cite{Lee2024}.
Such an interface will facilitate the construction of realistic low-energy models from first-principles calculations and their subsequent treatment within the ghost embedding framework. Extending GEM toward charge self-consistency is also appealing for material applications, such as systematic studies on competing phases across structural or stoichiometric phase spaces.

\section{Conclusions}
\label{sec:conclusions}

In this work, we have presented the Ghost Embedding Method package (GEM), an open-source implementation of the ghost-Gutzwiller approximation in its quantum-embedding formulation, for both zero- and finite-temperature calculations.
GEM provides a practical implementation of the ghost-embedding functional and its self-consistency conditions, formulated in terms of one-body density matrices rather than frequency-dependent Green's functions as in DMFT.
The number of auxiliary bath modes provides a systematic way of enlarging the variational space, connecting the Gutzwiller approximation at small bath size to DMFT in the infinite bath limit.
The software is organized around separate lattice, fragment, and impurity-solver components. This modular structure separates the ghost-embedding self-consistency from the solution of the interacting embedding problem and facilitates the implementation and testing of alternative impurity solvers. GEM includes an exact-diagonalization solver and provides a common interface through which external solvers can be incorporated.
We have illustrated the use of GEM through three representative applications. These examples demonstrate finite-temperature calculations of thermodynamic observables for the single-orbital Hubbard model, the determination of a finite-temperature magnetic phase diagram using multiple correlated fragments, and a zero-temperature calculation of a multi-orbital Hubbard-Kanamori model. Together, they illustrate the applicability of the same framework to thermodynamic properties, symmetry-broken phases, and multi-orbital correlated systems.
By providing the ghost-embedding approach as a documented and reusable software package, GEM makes this intermediate level of approximation between inexpensive static Gutzwiller methods and fully dynamical DMFT calculations accessible for model calculations, benchmarking, and methodological developments.
We expect its modular solver interface and compatibility with the broader scientific Python ecosystem to provide a basis for further developments and applications of ghost-embedding methods.

\section*{Acknowledgments}

We acknowledge helpful discussions with N. Hogan, G. Bellomia, C. Mejuto-Zaera, E. Zavatti, P. Filice, F. Kugler, A.M. Tagliente and M. Capone at various stages of development.


\paragraph{Funding information}
The Flatiron Institute is a division of the Simons Foundation.
N.L. gratefully acknowledges funding from the National Science Foundation under Award No. DMR-2532771 and from the Simons Foundation (Grant No. 00024037). 
Work by Y.Y. was supported by the U.S. Department of Energy (DOE), Office of Science, Basic Energy Sciences, Materials Science and Engineering Division, including the grant of computer time at the National Energy Research Scientific Computing Center (NERSC) in Berkeley, California. This part of research was performed at the Ames National Laboratory, which is operated for the U.S. DOE by Iowa State University under Contract No. DE-AC02-07CH11358. T.-H. L. gratefully acknowledges funding from the National Science and Technology Council (NSTC) of Taiwan under Grant Nos. NSTC 112-2112-M-194-007-MY3 and 115-2112-M-194-005, as well as support from the National Center for Theoretical Sciences (NCTS) in Taiwan.


\begin{appendix}
\numberwithin{equation}{section}

\section{Density-matrix fitting}
\label{app:denmat_fit}

The self-consistency conditions of the ghost-embedding functional Eq.~\eqref{eq:Lbeta} are given by the saddle-point equations Eqs.~(\ref{eq:saddle_R}-\ref{eq:saddle_lambda_c}). These are expressed as matching conditions of one-body density matrices belonging to the three auxiliary problems: the \textit{quasiparticle} problem with Hamiltonian given by Eq.~\eqref{eq:Hqp}, the \textit{interacting embedding} problem with Hamiltonian given by Eq.~\eqref{eq:Hemb} and the \textit{non-interacting embedding} problem with Hamiltonian given by Eq.~\eqref{eq:H0emb}.

At zero temperature, the ghost-Gutzwiller stationarity conditions can be solved in closed form, as shown in Ref.~\cite{Giuli2026}.
The auxiliary state is a Slater determinant, and the mixed
expectation values entering the update of the variational matrices are algebraic functions of the quasiparticle density matrix, involving only the matrix square
root $[\Delta(1-\Delta)]^{1/2}$, where $\Delta_{\alpha \beta}=\langle f^\dagger_{\alpha } f_{\beta} \rangle=\langle b_\beta b^\dagger_\alpha \rangle $.

At finite temperature this algebraic structure is lost.
GEM therefore imposes the finite-temperature conditions directly, as a pair of nonlinear least-squares problems in which a \emph{thermal} one-body density matrix is fitted to a set of target blocks.
This appendix documents that procedure, implemented in \texttt{python/gem/delta\_fit.py} and driven by \texttt{Fragment.update\_self\_energy} and
\texttt{Fragment.update\_hybridization}.

We consider a single fragment with $\nu$ physical (impurity) orbitals and $B \nu$ auxiliary (ghost) levels, where $B$ is the number of ghost levels per physical orbital.

\paragraph{The \texttt{Fragment.update\_self\_energy} routine.}

The fitting of $\Lambda_i$ and $\mathcal R_i$ is performed to satisfy the saddle point condition of $\Lambda^c_i$, Eq.~\eqref{eq:saddle_lambda_c}, and of $\mathcal D_i$, Eq.~\eqref{eq:saddle_D}. 
The fit of the one-body density matrix is performed by the \texttt{Fragment.update\_self\_energy} routine.
Both conditions involve the one-body density matrix of the non-interacting embedding problem defined in Eq.~\eqref{eq:H0emb}.
Denoting with ${H}^{0,\mathrm{emb}}_i$ the $2B\nu\times2B\nu$ single-particle matrix representation of the quadratic embedding Hamiltonian
$\hat H^{0,\mathrm{emb}}_i$, the corresponding one-body density matrix at temperature $T=1/\beta$ is
\begin{equation}
  \mathcal{F}_i(\Lambda_i,\mathcal R_i;\Lambda^c_i,\mathcal D_i)
  \;=\;
  \left[\,f_\beta\big({H}^{0,\rm emb}_i\big)\,\right]^{T},
  \qquad
  f_\beta(x) = \frac{1}{1+e^{\beta x}},
  \label{eq:F0emb}
\end{equation}
where the matrix function is evaluated by diagonalizing
${H}^{0,\mathrm{emb}}_i$. the transpose enforces the convention
$[\mathcal{F}_i]_{bc}=\langle a^\dagger_{ib} a_{ic} \rangle$ (where $a$ can either be an $f$ or a $b$ electron) used
throughout GEM. We denote by $\mathcal{F}^{11}_i$, $\mathcal{F}^{12}_i$ and $\mathcal{F}^{22}_i$ its $B\nu\times B\nu$ blocks, where $1$ labels the $f$ fermions (quasiparticle) sector, in which $\Lambda_i$ acts, and $2$ the $b$ fermions (embedding-bath) sector, in which $-\Lambda^c_i$ acts.
All the dependence of the saddle-point conditions Eqs.~(\ref{eq:saddle_D},~\ref{eq:saddle_lambda_c}) on $\Lambda_i$ and $\mathcal R_i$ is contained in Eq.~\eqref{eq:F0emb}.
The left-hand sides are the corresponding blocks $\langle b^\dagger_{i a} b_{i b} \rangle_{\rm emb}$ and $\langle c^\dagger_{i \alpha} b_{i b} \rangle_{\rm emb}$ of the one-body density matrix of the \emph{interacting} embedding problem Eq.~\eqref{eq:Hemb}, which are computed when calling \texttt{Fragment.solve\_impurity}. They are held fixed at the values returned by the impurity solver during the fit.

The routine therefore minimizes the cost function
\begin{equation}
  \chi^2_{i,\Sigma}(\Lambda_i,\mathcal R_i)
  = \sum_{a \le a'}
    \Big|\big[\mathcal{F}^{22}_i]_{a a^\prime}-\langle b^\dagger_{i a} b_{i a^\prime} \rangle_{\rm emb} \Big|^2
  + \sum_{a,\alpha}
    \Big|\big[\mathcal R^{T}_i\,\mathcal{F}^{12}_i \big]_{\alpha a} - \langle c^\dagger_{i \alpha} b_{ia} \rangle_{\rm emb} \Big|^2
  + \alpha_{\rm p}\,\big\|\,\mathsf{y}-\mathsf{y}_0\big\|^2 ,
  \label{eq:chi2_LR}
\end{equation}
where the first sum runs over the independent components of a Hermitian matrix and $\mathsf{y}$  is an array that collects all the independent real parameters of $\Lambda_i$ and $\mathcal R_i$, with $\mathsf{y}_0$ their value at the previous iteration of the self-consistency cycle.
Discarding for a moment the last term, the two residual blocks of Eq.~\eqref{eq:chi2_LR} contain $(B\nu)^2$ and $2B\nu^2$ independent real equations respectively, exactly as many as the free real parameters of a Hermitian $\Lambda_i$ and of a complex $\mathcal R_i$: the fit formally is a square nonlinear root-finding problem, which is however solved as a least-squares problem for the reasons given below.

The minimization of the distance is carried out with the Gauss-Newton algorithm implemented in \texttt{scipy.optimize.least\_squares}, using the analytic Jacobian of the residuals.
The only non-trivial ingredient of the latter is the derivative of the matrix Fermi function Eq.~\eqref{eq:F0emb}, which is not the naive derivative of
$f_\beta$, but rather its Fr\'echet derivative~\cite{WikipediaFrechetDerivative}, evaluated in the spectral (Daleckii--Kre\u{\i}n) representation~\cite[Eq.~(2.11) and Cor.~2.9]{Carlsson2019DaleskiiKrein}
\begin{equation}
  \delta\mathcal{F}_i
  = \Big[\,U\big(G\circ U^\dagger\,\delta {H}^{0,\rm emb}_i\,U\big)U^\dagger\Big]^{T},
  \qquad
  G_{mn} = \frac{f_\beta(\varepsilon_m)-f_\beta(\varepsilon_n)}{\varepsilon_m-\varepsilon_n},
  \label{eq:frechet}
\end{equation}
with ${H}^{0,\rm emb}_i = U\,{\rm diag}(\varepsilon)\,U^\dagger$, $G_{mn}\to f'_\beta(\varepsilon_m)$ in the degenerate
limit $\varepsilon_m\to\varepsilon_n$ and $\circ$
the Hadamard (element-wise) product. The cost
of the fit is dominated by the diagonalization of $\mathcal{H}^{0,\rm emb}_i$ that is a quadratic Hamiltonian and
is therefore negligible compared to a single solution of the interacting embedding problem.

The last term of Eq.~\eqref{eq:chi2_LR} is a proximal penalty that keeps the solution close to the previous iteration. It is needed, because the residuals are invariant under the gauge transformation $\Lambda_i \to u\,\Lambda_i\,u^\dagger$, $\mathcal R_i \to u\,\mathcal R_i$, with $u$ any unitary transformation in the space of $B\nu \times B\nu$ matrices, which reflects the freedom
in the choice of the ghost basis.

A second degeneracy appears at low temperature, where the
division of differences in $G_mn$ of Eq.~\eqref{eq:frechet} are exponentially small for ghost
levels far from the chemical potential, and therefore the corresponding directions become
numerically flat.
The penalty regularizes both, shifting all the singular values
of the augmented Jacobian by at least $\sqrt{2\alpha_{\rm p}}$, and selects among
the gauge-equivalent solutions the one closest to $\mathsf{y}_0$, which keeps the
ordering and the phases of the ghost levels continuous along the
self-consistency cycle and along parameter scans.
It does not bias the converged
result: at a fixed point of the cycle, $\mathsf{y}=\mathsf{y}_0$, and the penalty
vanishes together with its gradient. In practice, $\alpha_{\rm p}$ is taken in the
range $10^{-8}$--$10^{-6}$ (argument \texttt{move\_pen}, default $10^{-6}$) and
can be increased or decreased adaptively if a given step fails to converge.

Finally, the fit is performed on a real parametrization of $(\Lambda_i,\mathcal
R_i)$, obtained by packing into a single real vector $\mathsf{y}$ the real upper
triangle of $\Lambda_i$ including the diagonal, its strict imaginary upper
triangle, and the real and imaginary parts of $\mathcal R_i$ (routines
\texttt{pack\_params} and \texttt{unpack\_params} in \texttt{delta\_fit.py}).
Hermiticity of $\Lambda_i$ is thus imposed by construction rather than as a constraint, and the least-squares solver operates on unconstrained real variables.
When $S_z$ is a good quantum number, the whole procedure is applied independently to each spin block (\texttt{use\_Sz=True}), which halves the total number of fitted parameters and reduces the cost of the fitting.

\paragraph{The \texttt{Fragment.update\_hybridization} routine.}

The fitting of $\Lambda^c_i$ and $\mathcal D_i$ is performed to satisfy the saddle point condition of $\Lambda_i$ and $\mathcal R_i$,  provided by Eqs.(~\ref{eq:saddle_lambda},~\ref{eq:saddle_R}). They closely mimic the fit of $\Lambda_i$ and $\mathcal R_i$ presented in the previous paragraph, but in this case, the fit of the one-body density matrix is performed by the \texttt{Fragment.update\_hybridization} routine.

Both conditions involve the one-body density matrix of the non-interacting embedding problem given by Eq.~\eqref{eq:H0emb}, as previously shown in Eq.~\eqref{eq:F0emb}, and again
all the dependence of the saddle-point conditions on $\Lambda^c_i$ and $\mathcal D_i$, given by Eqs.~(~\ref{eq:saddle_R},~\ref{eq:saddle_lambda}), is contained in Eq.~\eqref{eq:F0emb}.
The left-hand sides are the corresponding blocks $\langle f^\dagger_{i a} f_{i b} \rangle_{\text{qp}}$ and $\sum_{j=1}^{\mathcal{N}} \sum_{\beta=1}^{\nu_j} \sum_{b=1}^{B \nu_j} [t^{ij}]_{\alpha \beta} [\mathcal{R}^\dagger_j ]_{\beta b} \langle f^\dagger_{ia} f_{j b} \rangle_{\text{qp}}$ of the one-body density matrix of the \textit{quasiparticle} problem Eq.~\eqref{eq:Hqp}, which are computed when calling \texttt{Lattice.solve\_qp}, are stored inside each fragment object, and are held fixed during the fit.
For simplicity we will call:
\begin{equation}
    \sum_{j=1}^{\mathcal{N}} \sum_{\beta=1}^{\nu_j} \sum_{b=1}^{B \nu_j} [t^{ij}]_{\alpha \beta} [\mathcal{R}^\dagger_j ]_{\beta b} \langle f^\dagger_{ia} f_{j b} \rangle_{\text{qp}} = [\Gamma_{i}^{\rm qp}]_{a \alpha}.
\end{equation}

The routine therefore minimizes the cost function
\begin{equation}
  \chi^2_{i,\Delta}(\Lambda^c_i,\mathcal D_i)
  = \sum_{ a \le a'}
    \Big|\big[\mathcal{F}^{11}_i]_{a a^\prime}-\langle f^\dagger_{i a} f_{i a^\prime} \rangle_{\rm qp} \Big|^2
  + \sum_{a,\alpha}
    \Big|\big[\mathcal{F}^{12}_i \, \mathcal D_i \big]_{a \alpha} - [\Gamma_i^{\rm qp}]_{a \alpha} \Big|^2
  + \alpha_{\rm p}\,\big\|\,\mathsf{y}-\mathsf{y}_0\big\|^2 ,
  \label{eq:chi2_LcD}
\end{equation}
where the first sum runs over the independent components of a Hermitian matrix and $\mathsf{y}$ is an array that collects all the independent real parameters of $\Lambda^c_i$ and $\mathcal D_i$, with $\mathsf{y}_0$ their value at the previous iteration of the self-consistency cycle.
Discarding for a moment the last term, the two residual blocks of Eq.~\eqref{eq:chi2_LcD} contain $(B\nu)^2$ and $2B\nu^2$ independent real equations respectively, exactly as many as the free real parameters of a Hermitian $\Lambda^c_i$ and of a complex $\mathcal D_i$: the fit formally is a square nonlinear root-finding problem, which is however solved as a least-squares problem.

The minimization of the distance is carried out with the Gauss-Newton algorithm implemented in \texttt{scipy.optimize.least\_squares} using the analytic Jacobian of the residuals and, similarly to \texttt{Fragment.update\_self\_energy}, the derivative of the matrix Fermi function Eq.~\eqref{eq:F0emb} is evaluated as a Fr\'echet derivative.
The cost of the fit is once again dominated by the diagonalization of $\mathcal{H}^{0,\rm emb}_i$ and
is negligible compared to a single solution of the interacting embedding problem.

Similarly to the previous method, the last term of Eq.~\eqref{eq:chi2_LcD} is a proximal penalty that keeps the solution close to the previous iteration. 
It keeps the ordering and the phases of the ghost levels continuous along the self-consistency cycle and along parameter scans.
It does not bias the converged result since the penalty
vanishes together with its gradient if the saddle point equations are satisfied.
In practice, $\alpha_{\rm p}$ is taken in the range $10^{-8}$--$10^{-6}$ (argument \texttt{move\_pen}, default $10^{-6}$) and can be increased or decreased adaptively, if a given step fails to converge.

Finally, the fit is performed on a real parametrization of $(\Lambda^c_i,\mathcal
D_i)$, obtained by packing them into a single real vector $\mathsf{y}$: the real upper triangle of $\Lambda^c_i$ including the diagonal, its strict imaginary upper triangle, and the real and imaginary parts of $\mathcal D_i$ (routines \texttt{pack\_params} and \texttt{unpack\_params}).
Hermiticity of $\Lambda^c_i$ is thus imposed by construction rather than as a constraint, and the least-squares solver operates on unconstrained real variables.
When $S_z$ is a good quantum number the whole procedure is applied independently to each spin block (\texttt{use\_Sz=True}), which halves the total number of fitted parameters and reduces the cost of the fitting.

\end{appendix}





\bibliography{biblio}


\end{document}